\documentclass[a4paper]{jpconf}
\usepackage{graphicx}

\usepackage{citesort}
\usepackage{xspace}
\usepackage{amssymb,amsmath}
\usepackage{upgreek}
\usepackage{color}

\newcommand{\ptt}{\ensuremath{p_{\mathrm{T}}}\xspace}
\newcommand{\pb}{\mbox{Pb--Pb}\xspace}
\newcommand{\gvc}{\mbox{\rm GeV$\kern-0.15em /\kern-0.12em c$}\xspace}
\newcommand{\gvcc}{\mbox{\rm GeV$\kern-0.15em /\kern-0.12em c^2$}\xspace}
\newcommand{\mvc}{\mbox{\rm MeV$\kern-0.15em /\kern-0.12em c$}\xspace}
\newcommand{\mvcc}{\mbox{\rm MeV$\kern-0.15em /\kern-0.12em c^2$}\xspace}
\newcommand{\ph}{\ensuremath{\phi}\xspace}
\newcommand{\phm}{\ensuremath{\phi(1020)}\xspace}
\newcommand{\ks}{\ensuremath{\mathrm{K^{*0}}}\xspace}
\newcommand{\dndy}{\ensuremath{\mathrm{d}N\kern-0.15em /\kern-0.12em\mathrm{d}y}\xspace}
\newcommand{\npart}{\ensuremath{\langle N_{\mathrm{part}}\rangle}\xspace}
\newcommand{\mpt}{\ensuremath{\langle p_{\mathrm{T}}\rangle}\xspace}
\newcommand{\stwo}{\mbox{\ensuremath{\sqrt{s_{\mathrm{NN}}}=2.76\;\mathrm{TeV}}}\xspace}
\newcommand{\ssvn}{\mbox{\ensuremath{\sqrt{s}=7\;\mathrm{TeV}}}\xspace}
\newcommand{\Dphi}{\ensuremath{\mathit{\Delta}\varphi}\xspace}

\begin{document}
\title{Hadronic resonance production in Pb--Pb collisions at the ALICE experiment}

\author{A G Knospe (for the ALICE Collaboration)}

\address{Department of Physics, The University of Texas at Austin, 1 University Station C1600, Austin, Texas, USA 78712-0264}

\ead{anders.knospe@cern.ch}

\begin{abstract}
Measurements of the yields of hadronic resonances (relative to non-resonances) in high-energy heavy-ion collisions allow the chemical freeze-out temperature and the time between chemical and thermal freeze-out of the collision system to be studied, while modifications to resonance masses and widths could be a signature of chiral symmetry restoration.  The spectra (for \ptt$<$ 5 \gvc), total integrated yields, ratios to non-resonances ($\phi/\pi$ and $\phi$/K), mass, and width of the \phm meson and the uncorrected yields, mass, and width of the \mbox{K$^{*}(892)^{0}$} and \mbox{$\overline{\mathrm{K}^{*}}(892)^{0}$} mesons have been measured using the ALICE detector for \pb collisions at \stwo.  These measurements will be compared to results from other collision systems and energies.  Angular correlations between leading trigger hadrons and \phm mesons have been measured in \pb and pp collisions; the mass and width of the \phm meson as a function of the correlation angle will be presented.

\end{abstract}

\section{Introduction}

The production of resonances occurs both during the transition (at a critical temperature of \mbox{$\approx$~160 MeV}~\cite{Aoki_TC1,Aoki_TC2,Borsanyi_TC}) from the quark-gluon plasma (QGP) to the hadronic phase and in the hadronic phase itself due to regeneration~\cite{Bleicher_Stoecker,Markert_thermal,Vogel_Bleicher}.  Due to their short lifetimes (a few fm$\kern-0.15em /\kern-0.12em c$), many resonances will decay during the hadronic phase; their decay products will tend to be re-scattered in the hadronic medium, making experimental reconstruction of the resonance impossible~\cite{Bliecher_Aichelin}.  The temperature evolution and lifetime of the hadronic phase affect the relative strengths of resonance-generating processes and re-scattering, and therefore ratios of resonance yields to non-resonance yields~\cite{Torrieri_thermal}.  Particle ratios have been predicted as functions of the chemical freeze-out temperature and the elapsed time between chemical and thermal freeze-out using thermal models~\cite{Andronic2009,PBM2011,AndronicQM2011,Torrieri_thermal,Markert_thermal}.  In principle, measurements of two different particle ratios are needed to tune the thermal models and determine uniquely the chemical freeze-out temperature and the lifetime of the hadronic medium; in practice, many measured ratios are used to obtain a best-fit value for these model parameters.

It is expected that chiral symmetry is restored at around the same critical temperature as the deconfinement phase transition~\cite{Petreczky}, with the quark-antiquark condensate decreasing towards 0 with increasing temperature~\cite{Rapp2009}.  Resonances that interact with the medium in the mixed or early hadronic phases, when chiral symmetry is at least partially restored, may be shifted off of their mass shells and exhibit broader widths than observed in vacuum~\cite{Rapp2009,Brown_Rho,Brodsky_chiral}.  For example, it has been predicted that the $\rho$ meson should ``melt" (its width approaches its mass) near the critical temperature~\cite{Rapp2009}, with a mass shift of a few tens of MeV at most~\cite{Eletsky}.  However, the modified resonance spectral functions will only be observable if a large fraction of resonances decay while the medium is chirally restored.  UrQMD~\cite{UrQMD} calculations indicate that regeneration (which adds resonances with vacuum properties) and re-scattering effects (which reduce the resonance signal) will be strongest for $\ptt<2$~\gvc.  It was argued in~\cite{Markert_corr}, accounting for the time dependence of the QGP temperature and the time dilation of resonance decays, that experimental searches at the LHC for the signatures of chiral symmetry restoration should focus on resonances with $\ptt<10$~\gvc.  It has been predicted that the majority of reconstructable resonances with $\ptt>2$~\gvc will be \textit{produced} when the density of the medium is high~\cite{Vogel_Aichelin_Bliecher}; however, estimating the fraction of reconstructable resonances that \textit{decay} in the chirally restored medium is a more complicated question and further input is needed from the theoretical community~\cite{Markert_corr}.  Resonances with shorter lifetimes (a few fm/$c$) are more likely to exhibit mass shifts or width broadening, meaning that the $\mathrm{K}^{*}(892)^{0}$ would be more likely than the \ph(1020) to exhibit these signatures.



These proceedings present measurements of the \mbox{\phm}, \mbox{K$^{*}(892)^{0}$}, and \mbox{$\overline{\mathrm{K}^{*}}(892)^{0}$} resonances in \pb collisions at \stwo using the ALICE detector at the LHC.  For the sake of brevity, \mbox{\phm} will be hereafter be denoted simply by \ph.  The \mbox{K$^{*}(892)^{0}$} and \mbox{$\overline{\mathrm{K}^{*}}(892)^{0}$} mesons will be collectively referred to as \ks and all results will be for the combination of the particle and antiparticle.  The ALICE detector will be described in Section~\ref{sec:alice}.  The method used to extract the masses, widths, and yields of the resonances as functions of \ptt will be described in Section~\ref{sec:signal:extraction} and measurements of the \ph and \ks masses and widths as functions of \ptt will be presented and discussed in Section~\ref{sec:signal:mass_width}.  The spectra for \ptt$<$ 5 \gvc are presented is Section~\ref{sec:spectra}.  Resonance-to-non-resonance ratios and measurements of the mean transverse momentum of the \ph meson are presented and discussed in Section~\ref{sec:ratios_mpt}.  The method of jet-resonance (or hadron-resonance) correlations has been proposed as a way of preferentially selecting for resonances (with $\ptt\gtrsim 2\gvc$) that originated in the mixed or partonic phases, when chiral symmetry was still at least partially restored.  The hadron-resonance correlation method will be described in Section~\ref{sec:correlations}.

\section{The ALICE Detector}
\label{sec:alice}

The ALICE detector~\cite{ALICE_detector} is the only detector at the LHC dedicated primarily to the study of heavy-ion collisions.  It was designed to operate at high particle multiplicities, which were expected to be up to three orders of magnitude greater than for pp collisions at the same energy and 2-5 times greater than the multiplicities seen in heavy-ion collision at RHIC.  The ALICE detector provides extensive particle tracking and identification at mid-rapidity \mbox{$(|\eta|\lesssim 0.9)$} as well as muon tracking and identification at forward rapidity \mbox{$(-2.5>\eta>-4)$}.  The components of the ALICE detector most directly related to the results presented in this article will be described below.

The Inner Tracking System (ITS) is a silicon detector that surrounds the interaction point, with six layers between radii 3.9~cm and 43~cm from the beam axis.  The ITS is used to reconstruct the collision vertex and provide hits used in particle tracking.  Particle tracking is mainly provided by the large (90~m$^{3}$) cylindrical Time Projection Chamber (TPC)~\cite{ALICE_TPC}.  The TPC also allows particles to be identified through their energy loss.  The Time-of-Flight Detector (TOF) is an array of multi-gap resistive-plate chambers that sits outside the TPC and measures particle speed.  The TOF allows for kaons to be distinguished from pions for \mbox{$p_{\mathrm{T}}<2.5-3$ \gvc} and for kaons to be separated from protons for \mbox{$p_{\mathrm{T}}<3.5-4$ \gvc} (by better than 3$\sigma$ in both cases~\cite{ALICE_PPR_Vol1,ALICE_TOF}).  In the present analysis, TOF information was not used to identify the decay daughters of \ph and \ks, although it will be used in the future.  The V0 detector consists of two arrays of scintillator counters that sit at large pseudorapidities ($2.8<\eta<5.1$ and $-3.7<\eta<-1.7$) on opposite sides of the interaction point; it provides minimum-bias triggers for the central barrel detectors.

\section{Resonance Signals}
\label{sec:signal}

\subsection{Signal Extraction}
\label{sec:signal:extraction}

\begin{figure}
\begin{minipage}{19pc}
\includegraphics[width=19pc]{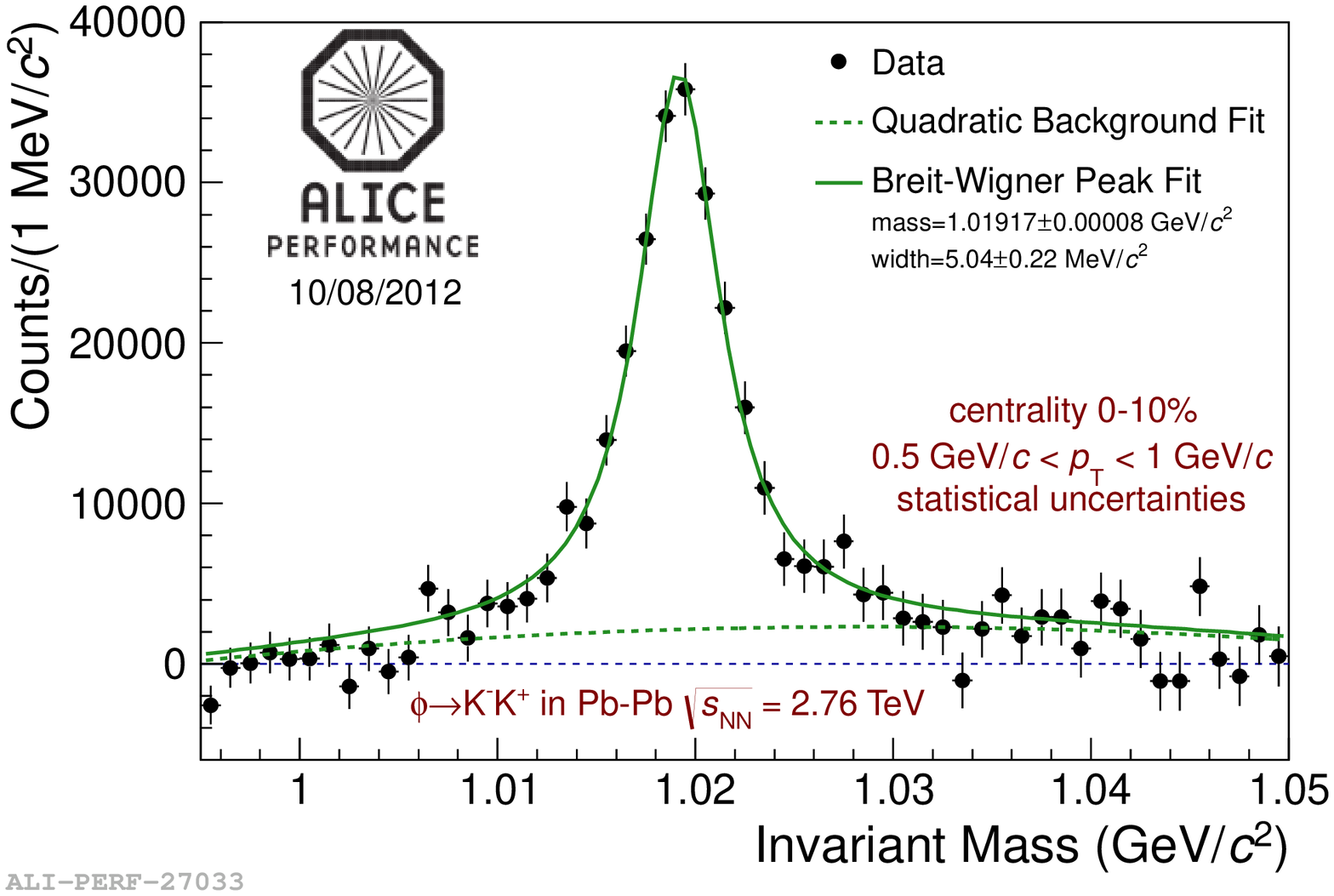}
\end{minipage}
\hspace{-15pc}
(a)
\hspace{13pc}
\begin{minipage}{18.5pc}
\includegraphics[width=19pc]{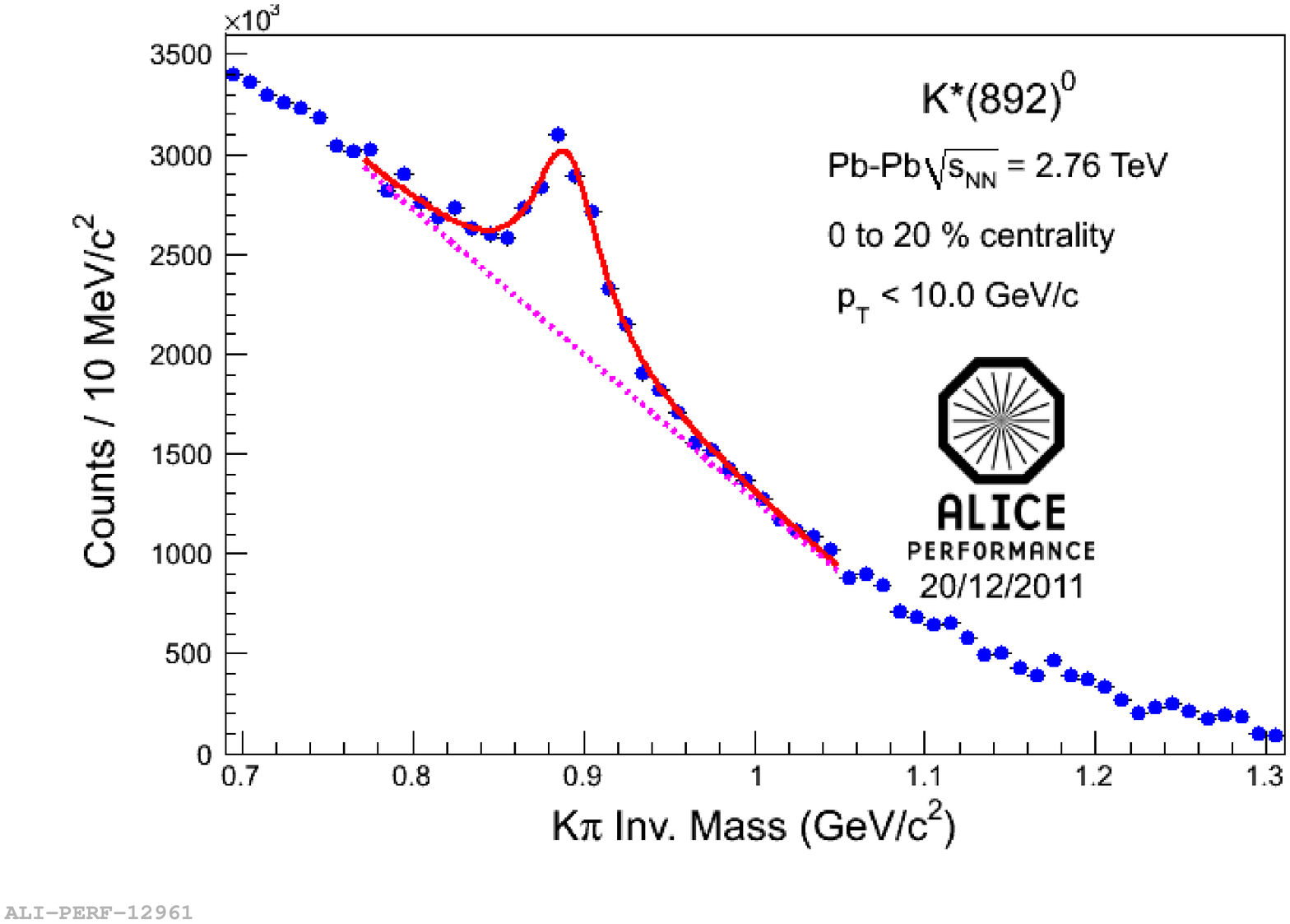}
\end{minipage}
\hspace{-15pc}
(b)
\caption{Invariant mass distributions after background subtraction of \ph (a) and \ks (b) mesons in central \pb collisions at \stwo.  For each particle, the residual background is described using a polynomial fit (dashed curves) and the peak is fit using Breit-Wigner function added to the residual background (solid curves).  For \ph: centrality 0-10\%, \mbox{0.5 \gvc$<\ptt<1$ \gvc}, second-order polynomial residual background.  For \ks: centrality 0-20\%, \mbox{\ptt$<10$\gvc}, first-order polynomial residual background.} 
\label{fig:signal:extraction:invmass_sub}
\end{figure}

The decays of \ph mesons to oppositely charged kaons (branching ratio = $0.489\pm 0.005$~\cite{PDG}) were reconstructed in 9.5 million \pb collisions.  The kaons were identified by applying a cut to their measured energy loss in the ALICE TPC.  The measured energy loss was required to deviate from the expected mean value by $<2\sigma_{\mathrm{K}}$, where $\sigma_{\mathrm{K}}$ is the energy-loss resolution ($<5\%$ for isolated tracks~\cite{ALICE_TPC}).  An invariant-mass distribution of pairs (with $|y_{\mathrm{pair}}|<0.5$) of oppositely charged kaons from the same event was generated.  In order to remove the large number of random combinations, combinatorial backgrounds were generated using two different methods.  The like-charge\footnote{also known as ``like-sign"} combinatorial background was generated by pairing each kaon candidate with kaons from the same event with the same charge.  The like-charge combinatorial background is $2\sqrt{n_{--}n_{++}}$, where $n_{--}$ and $n_{++}$ are the numbers of K$^{-}$K$^{-}$ and K$^{+}$K$^{+}$ pairs, respectively, in each invariant mass bin.  A second combinatorial background was generated by mixing oppositely charged kaon tracks from different events.  The pairs of mixed events were required to differ by \mbox{$<5$ cm} in $z$ vertex position and $<10\%$ in centrality percentile; each identified kaon track was mixed into $\sim 5$ other events.  The mixed-event combinatorial background was scaled so that its integral in a region surrounding the \ph peak\footnote{The exact boundaries of this normalization region ($B_{1}\;<\;m(\mathrm{KK})\;<\;B_{2}$ or $B_{3}\;<\;m(\mathrm{KK})\;<\;B_{4}$) were varied.  The following six sets of boundaries were used (units of \gvcc): $(B_{1},B_{2},B_{3},B_{4})$=(1,1.01,1.03,1.06), (1,1.006,1.04,1.06), (1,1.006,1.03,1.06), (1,1.01,1.035,1.06), (1,1.006,1.04,1.05), and (0.995,1.01,1.03,1.06).} was the same as the integral of the unlike-charge distribution in the same region.  The mixed-event combinatorial background was chosen as the primary method for this analysis due to its smaller statistical uncertainties and fluctuations.

After the subtraction of the combinatorial background, a \ph peak could be observed sitting on a residual background (See Figure~\ref{fig:signal:extraction:invmass_sub}a), which was present due to the fact that neither choice of combinatorial background could account for correlated sources of unlike-charge background.  A polynomial fit was performed on a region surrounding the peak in order to parametrize this residual background.  Second-order polynomials were used as the primary fitting functions.  To extract the \ph yield, the unlike-charge invariant mass distribution was integrated and the integral of the residual background fit was subtracted.  The unlike-charge distribution was also fit with a relativistic Breit-Wigner function added to the residual background fit.  This provided a second method for extracting the particle yield and also allowed the mass and width of the \ph meson to be extracted.  The systematic uncertainties in the \ph yield take into account variations in the yield due to different kaon identification (TPC energy loss) cuts, combinatorial backgrounds, normalization regions, residual background parameterizations\footnote{First- and third-order polynomial fits were also performed.}, residual background fit regions, and methods of extracting the yield from the peak (integrating the histogram vs. integrating the fit).  The systematic uncertainties also include contributions that arise from varying the track quality cuts, uncertainties in the amount of material used in simulations of the ALICE detector (in the efficiency calculation), and the uncertainty in centrality selection.

The extraction of the \ks yields, mass, and width was performed in a similar fashion (See Figure~\ref{fig:signal:extraction:invmass_sub}b).  The decays of \ks mesons to $\pi^{\pm}\mathrm{K}^{\mp}$ pairs (branching ratio = $0.66601\pm0.00006$~\cite{PDG}) were reconstructed in 3.4 million \pb collisions.  The kaons were again identified by applying a $2\sigma_{\mathrm{K}}$ cut on energy loss in the TPC, while the pions were identified by applying a $2\sigma_{\pi}$ cut on the energy loss.  As in the \ph analysis, both like-charge and mixed-event combinatorial backgrounds were constructed.  A first-order polynomial was used as the primary fit function for the residual background.

\subsection{Masses and Widths}
\label{sec:signal:mass_width}

\begin{figure}
\begin{minipage}{18.5pc}
\includegraphics[width=18.5pc]{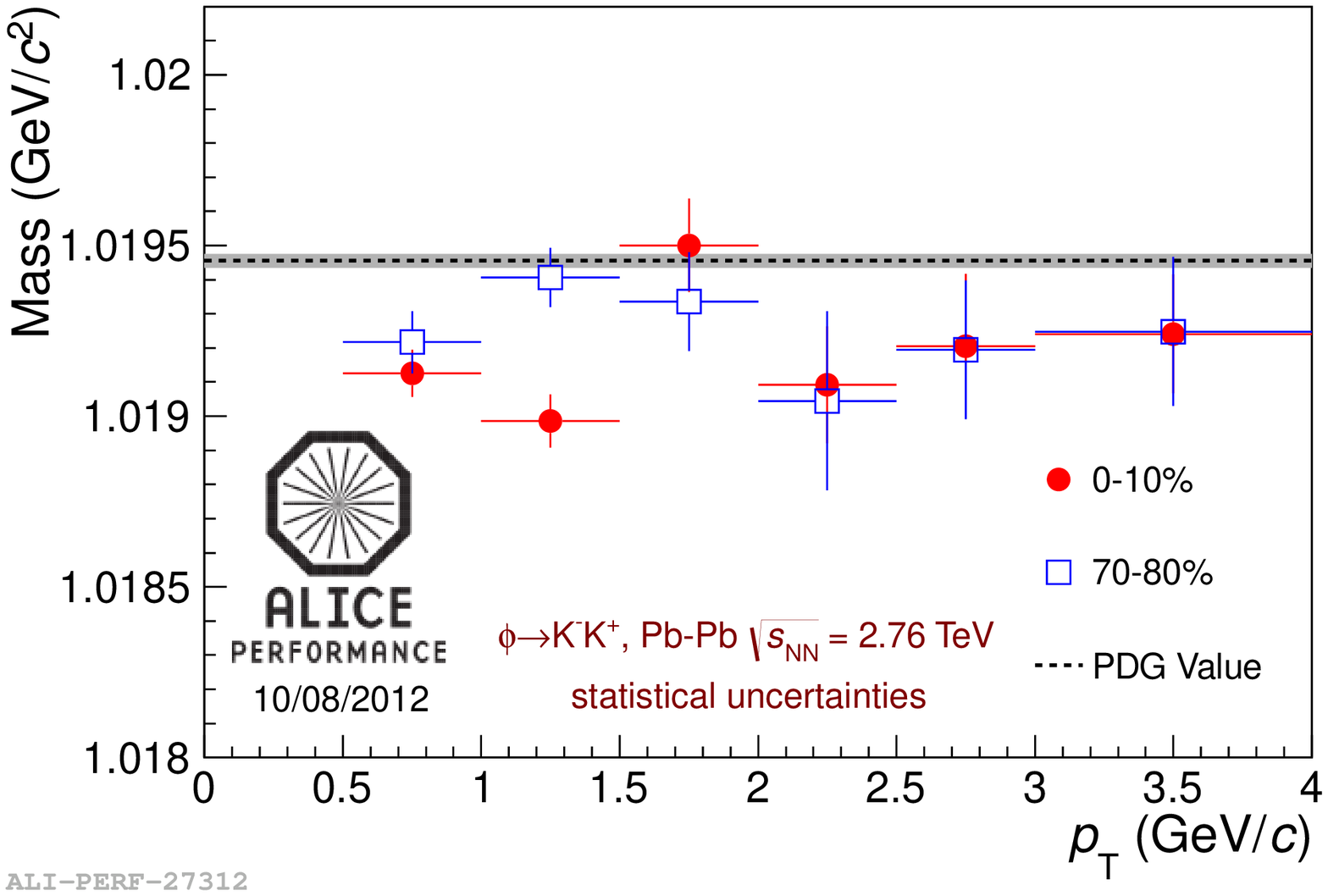}
\end{minipage}
\begin{minipage}{0pc}
\hspace{-16pc}\vspace{9pc}
(a)
\end{minipage}
\begin{minipage}{18.5pc}
\includegraphics[width=18.5pc]{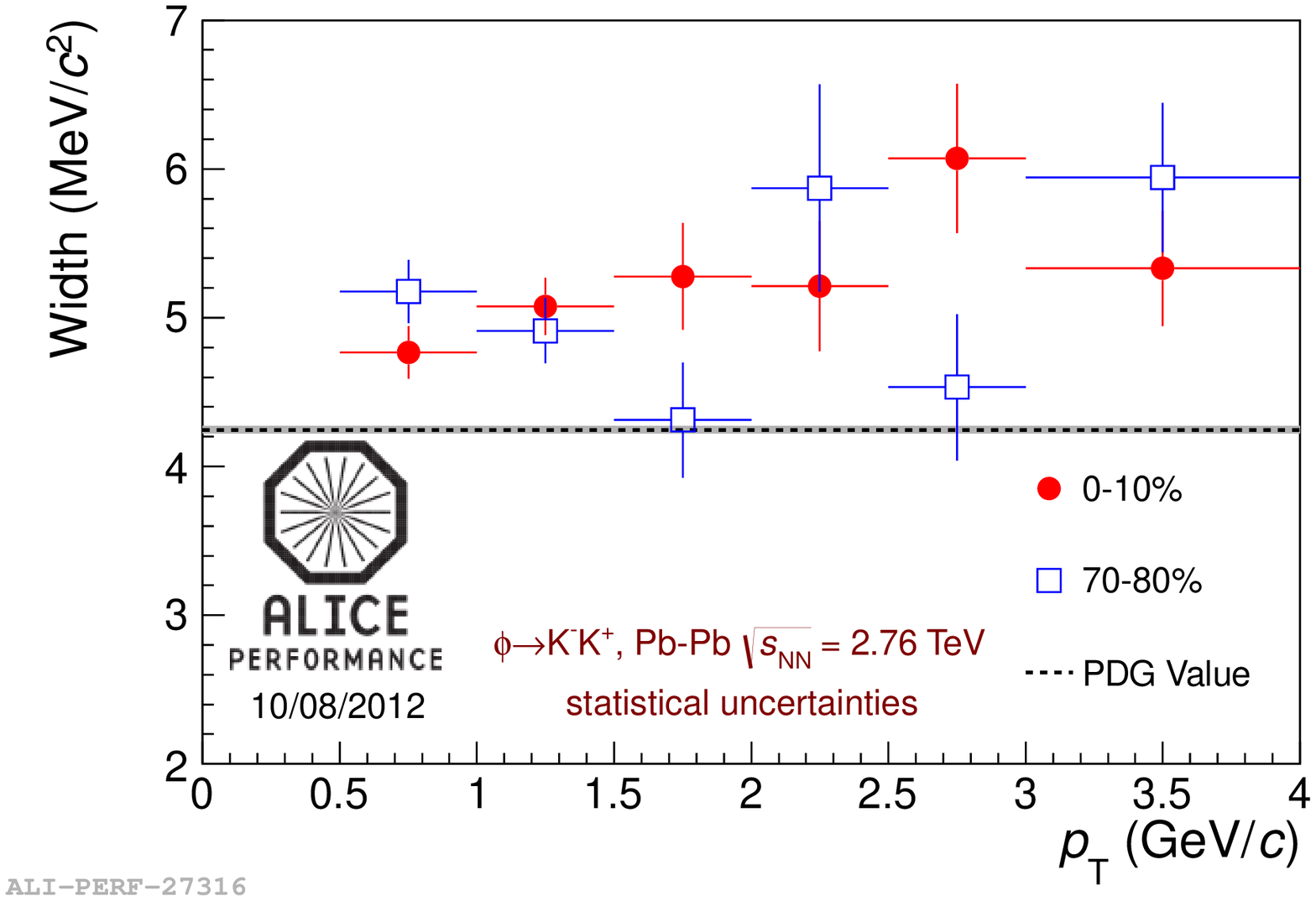}
\end{minipage}
\begin{minipage}{0pc}
\hspace{-16pc}\vspace{9pc}
(b)
\end{minipage}
\caption{Measured \ph meson mass (a) and width (b) in \pb collisions at \stwo.  Red circles: centrality 0-10\%; blue squares: centrality 70-80\%.  The dashed lines indicate the nominal values of the \ph mass and width~\cite{PDG}.} 
\label{fig:signal:mass_width:phi}

\begin{minipage}{18.5pc}
\includegraphics[width=18.5pc]{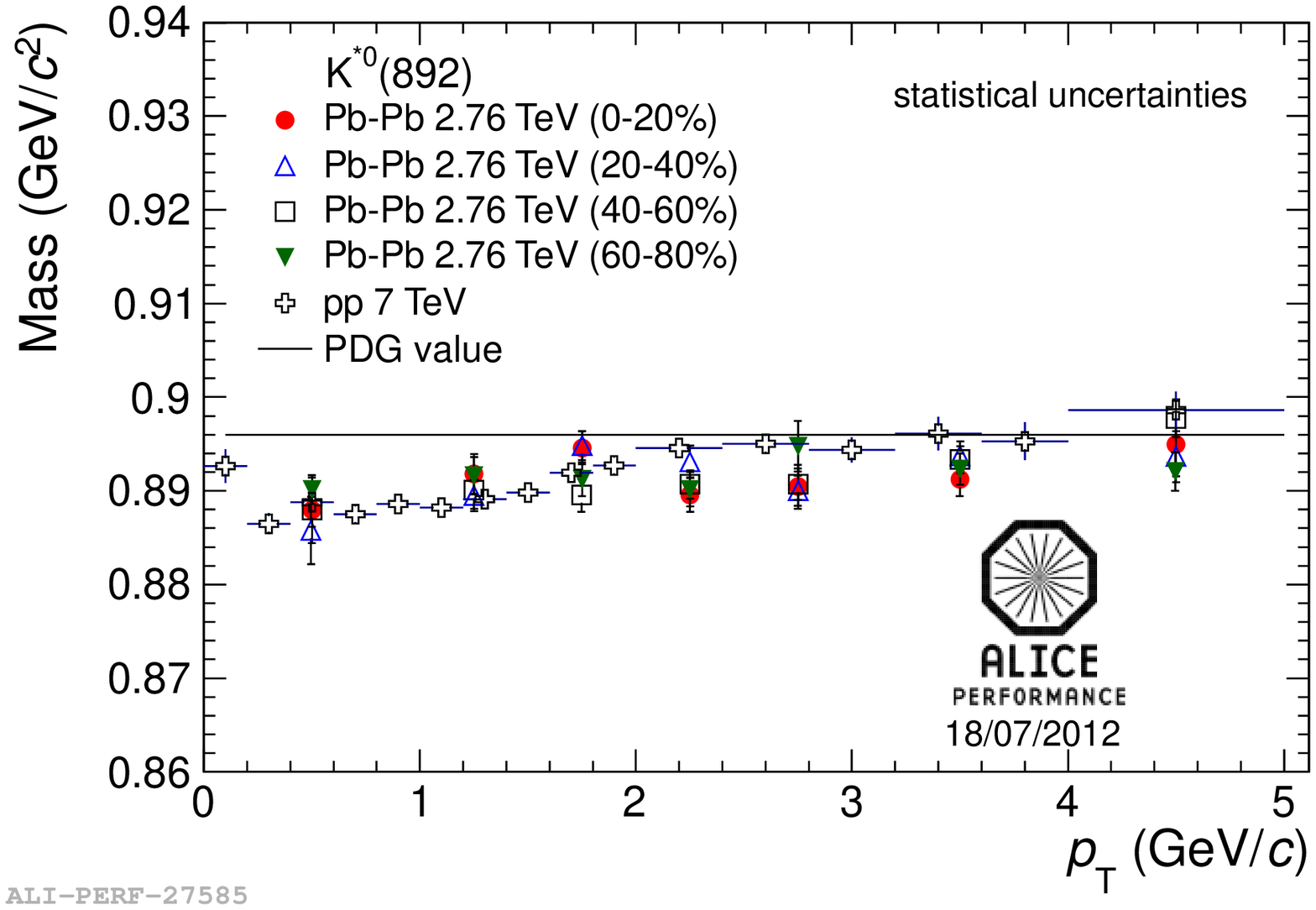}
\end{minipage}
\begin{minipage}{0pc}
\hspace{-16pc}\vspace{-6pc}
(a)
\end{minipage}
\begin{minipage}{18.5pc}
\includegraphics[width=18.5pc]{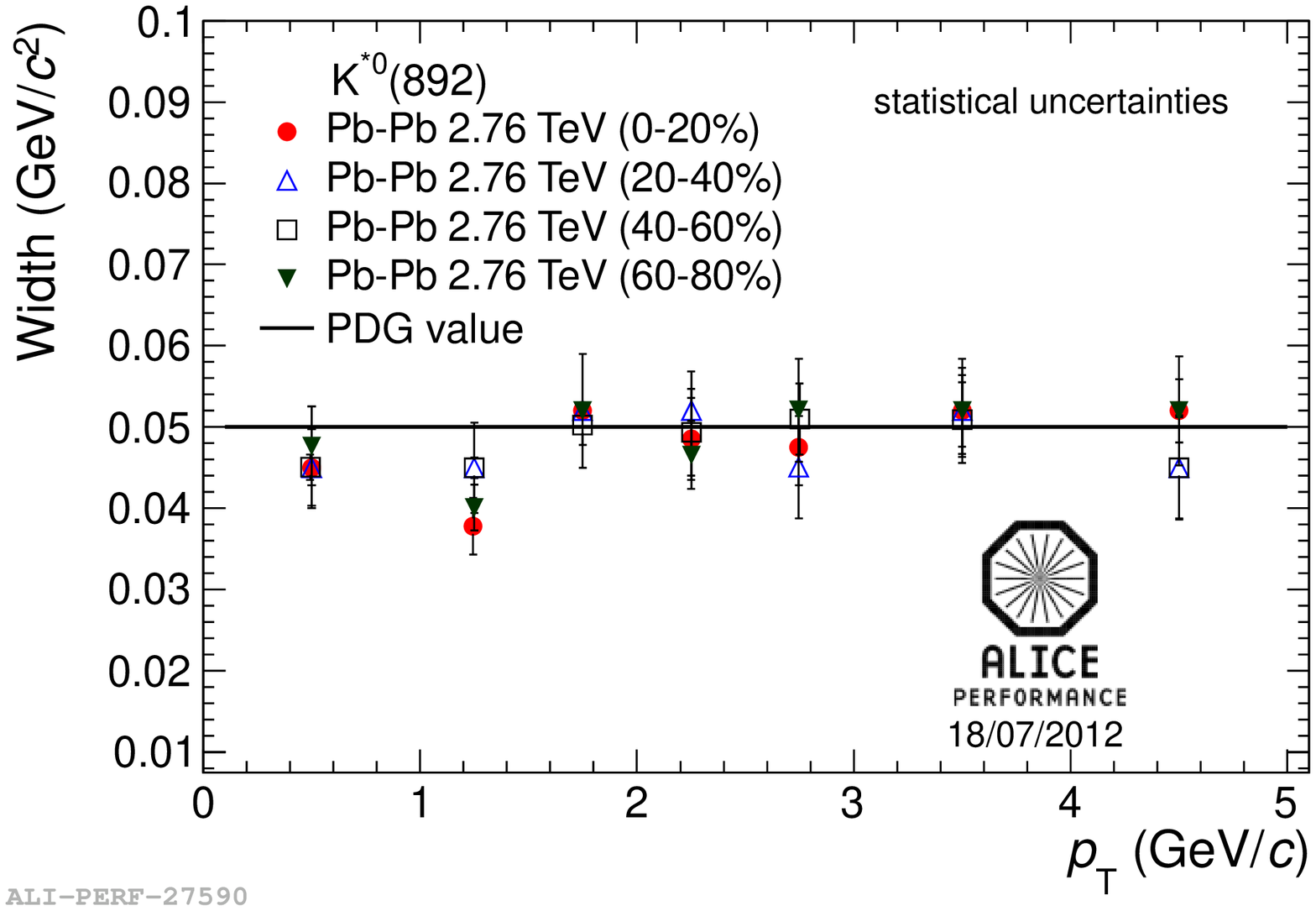}
\end{minipage}
\begin{minipage}{0pc}
\hspace{-16pc}\vspace{-6pc}
(b)
\end{minipage}
\caption{Measured \ks meson mass (a) and width (b) in \pb collisions at \stwo in four centrality bins, along with the results in pp collisions at \ssvn~\cite{Markert_SQM2011}.  The black lines indicate the nominal values of the \ks mass and width~\cite{PDG}.}
\label{fig:signal:mass_width:ks}
\end{figure}

The mass and width of \ph mesons are shown as functions of transverse momentum in Figure~\ref{fig:signal:mass_width:phi} for central and peripheral \pb collisions.  The measured values of the \ph mass are within 0.5~\mvcc of the nominal value, while the measured values of the width are within 2 \mvcc of the nominal value.  There is no obvious centrality dependence in either the mass or the width.  The mass and width of \ph mesons have also been extracted from simulated \pb collisions, with particle production via HIJING and the ALICE detector simulated using GEANT3.  The deviations from the nominal mass and width observed in these simulations\footnote{The values of the simulated mass are observed to be within 0.5~\mvcc of the vacuum value, while the values of the simulated width are observed to be within 1-2~\mvcc of the vacuum value.  Studies of these simulated \ph peaks are ongoing.} are similar to the deviations observed for the real data, suggesting that the shifts in the mass and width can be explained purely by detector effects.

The mass and width of \ks are shown as functions of transverse momentum in Figure~\ref{fig:signal:mass_width:ks} for \pb collisions at \stwo in multiple centrality bins.  The measured values of the \ks mass are consistent with the values measured in pp collisions at \ssvn, which suggests that the deviation from the nominal value is an instrumental effect\footnote{The reconstructed masses and widths of simulated \ks peaks are being studied in order to confirm whether or not these deviations are detector effects.}.  The measured values of the width are generally consistent with the nominal value.  As with the \ph meson, there is no apparent centrality dependence in the mass or width.

\section{Spectra}
\label{sec:spectra}

\begin{figure}
\includegraphics[width=18pc]{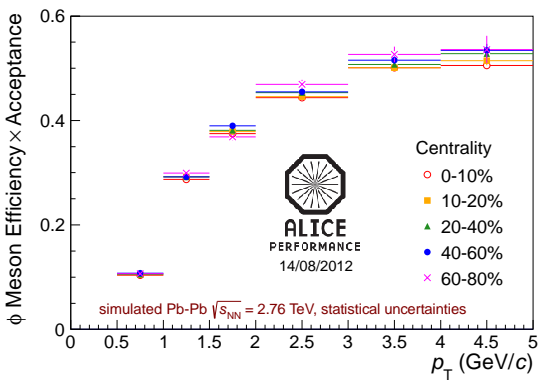}
\hspace{1pc}
\begin{minipage}[b]{14pc}
\caption{Efficiency $\times$ acceptance for the \ph meson, calculated from simulated \pb events.  See the text for discussion.}
\label{fig:spectra:efficiency}
\end{minipage}
\end{figure}

The efficiency $\times$ acceptance for the \ph meson was calculated from simulated 1.82 million \pb events; particles were generated using HIJING and their interactions with the ALICE detector were simulated using GEANT3.  The same event-selection and track-quality cuts were used in the analyses of the real and simulated data.  Figure~\ref{fig:spectra:efficiency} shows efficiency $\times$ acceptance for the \ph meson.  While the energy-loss followed a Gaussian distribution in the real data, this behavior was not reproduced in the simulation.  For this reason, a separate factor $(\varepsilon_{\mathrm{d}E/\mathrm{d}x})$ was applied to correct for the efficiency of the TPC energy-loss cuts used to identify kaons; for the default energy-loss cuts (a $2\sigma_{\mathrm{K}}$ cut) $\varepsilon_{\mathrm{d}E/\mathrm{d}x}=91\%$.  More simulated events must be analyzed in order for the efficiency calculation to be extended to higher transverse momenta; it is for this reason that the spectra reported in this article are only for $\ptt<5$~\gvc.

\begin{figure}
\begin{minipage}{18.5pc}
\includegraphics[width=18.5pc]{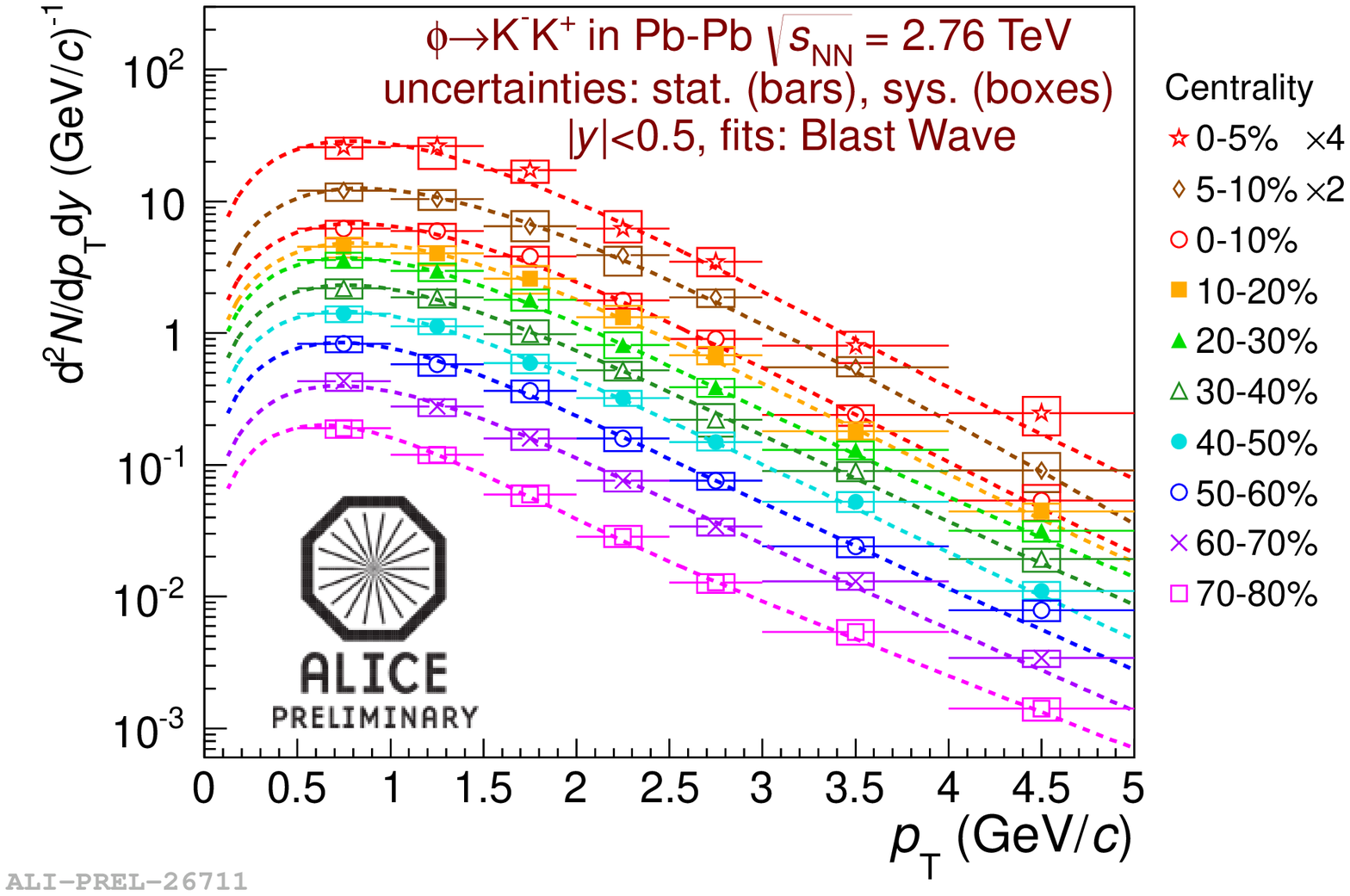}
\end{minipage}
\begin{minipage}{0pc}
\vspace{-10pc}
\hspace{-16pc}
(a)
\end{minipage}
\begin{minipage}{18pc}
\includegraphics[width=18.5pc]{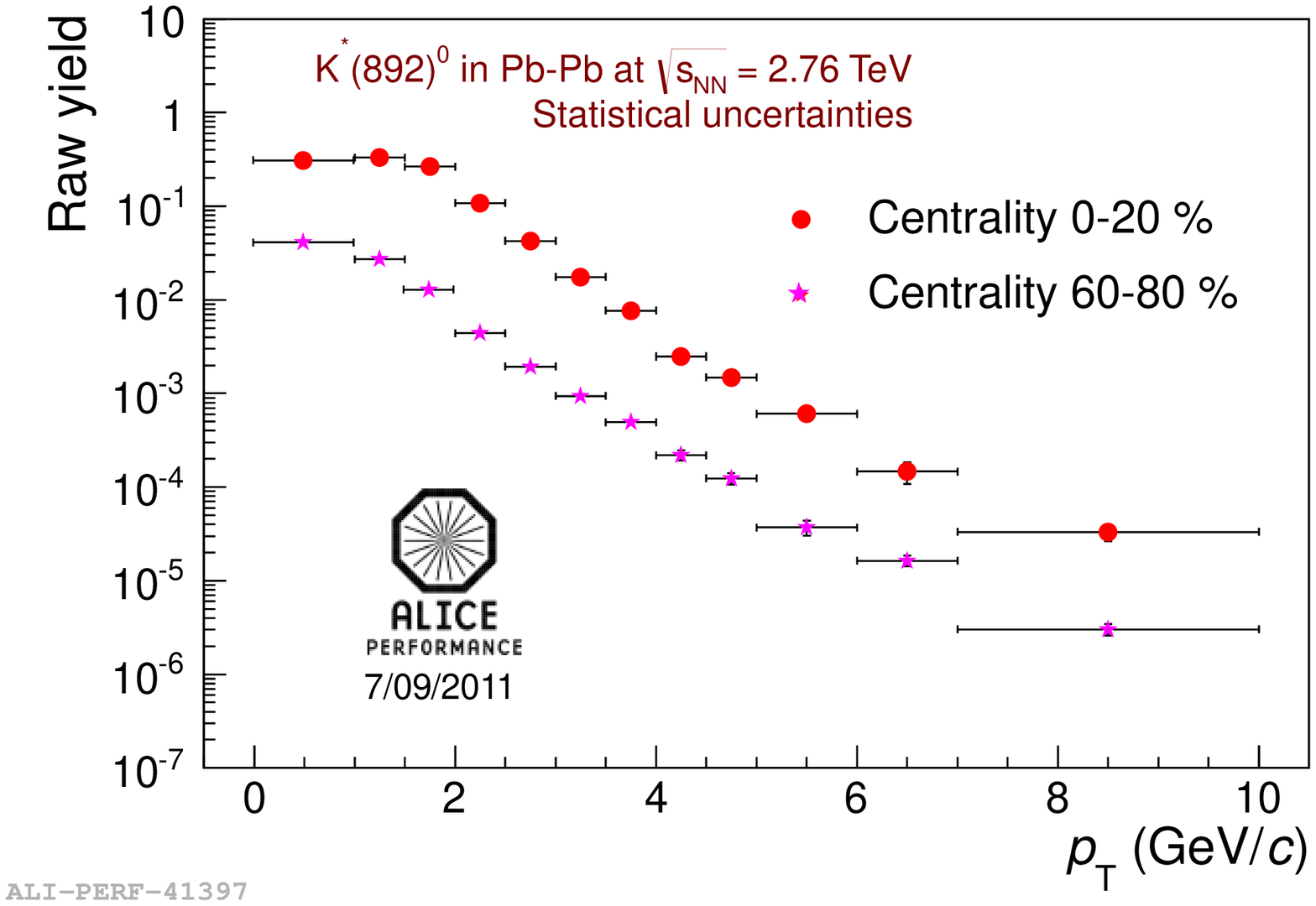}
\end{minipage}
\begin{minipage}{0pc}
\vspace{-10pc}
\hspace{-3pc}
(b)
\end{minipage}
\caption{Corrected spectra of \ph mesons (a) and uncorrected yields of \ks mesons (b) in \pb collisions at \stwo in multiple centrality bins.  Blast Wave fits to the \ph spectra are shown.  Statistical (systematic) uncertainties are represented by bars (boxes).} 
\label{fig:spectra:plots}
\end{figure}

Figure~\ref{fig:spectra:plots}a shows the corrected \ph spectra (d$^{2}N/$d$p_{\mathrm{T}}$d$y$) in multiple centrality bins; these spectra have been corrected for the efficiency $\times$ acceptance, energy-loss cut efficiency, and the branching ratios.  The \ph spectra were fit with Boltzmann-Gibbs Blast Wave functions~\cite{BoltzmannGibbsBlastWave}.  Figure~\ref{fig:spectra:plots}b shows the \textit{uncorrected} \ks yields; signals are observed for 
\ptt $<$ 10 \gvc.

\section{Ratios and Mean $\boldsymbol{p_{\mathrm{T}}}$}
\label{sec:ratios_mpt}

\begin{figure}[h]
\begin{minipage}{18.5pc}
\includegraphics[width=18.5pc]{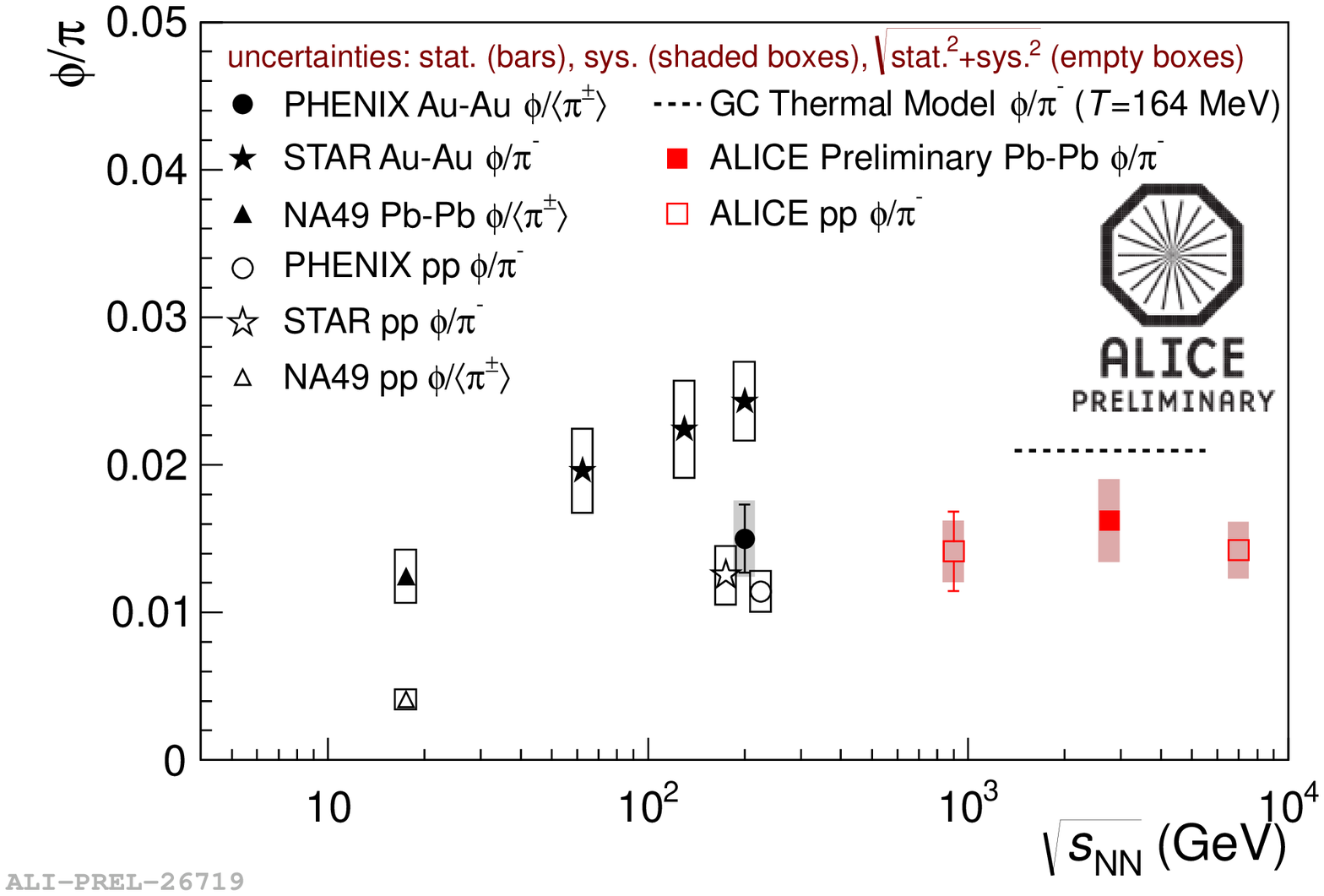}
\end{minipage}
\begin{minipage}{0pc}
\hspace{-15.5pc}\vspace{-3pc}
(a)
\end{minipage}
\begin{minipage}{18pc}
\includegraphics[width=18.5pc]{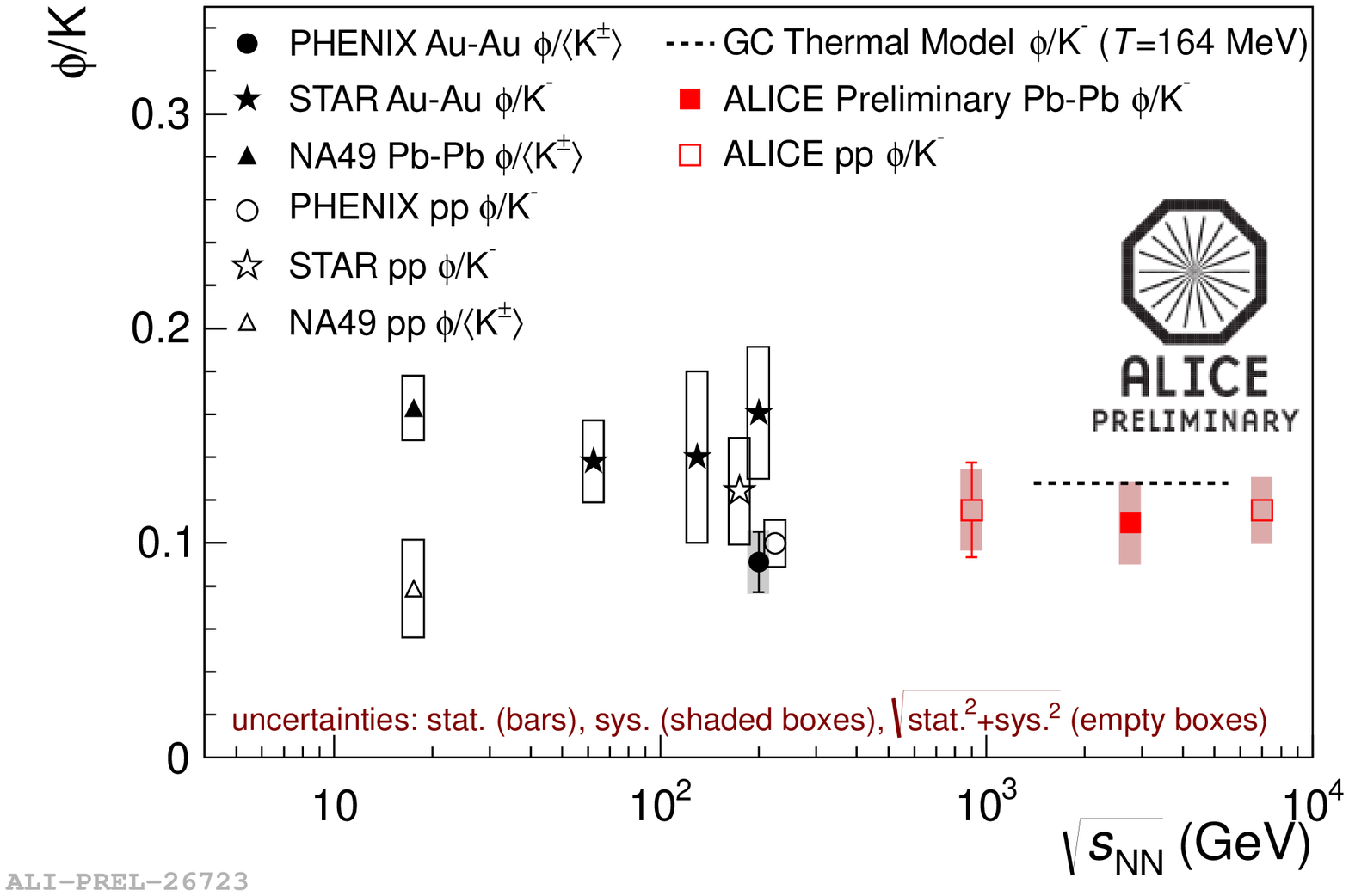}
\end{minipage}
\begin{minipage}{0pc}
\hspace{-15pc}\vspace{-3pc}
(b)
\end{minipage}
\caption{Ratios of resonance yields to non-resonance yields as functions of $\sqrt{s_{\mathrm{NN}}}$ from SPS to LHC energies.~\cite{NA49,STAR_phi_2009,PHENIX_piKp_pp_2011,PHENIX_mesons_pp_2011,PHENIX_phi_AuAu_2005,ALICE_piKp_900GeV,ALICE_strange_900GeV,ALICE_KstarPhi_7TeV}  ALICE \pb data are for centrality 0-5\%.  (a) shows the $\phi/\pi$ ratios.  (b) shows the $\phi/\mathrm{K}$ ratio.  The dashed lines indicate thermal model predictions~\cite{AndronicQM2011} for \pb collisions at \ssvn.} 
\label{fig:ratios_mpt:phi_ratios_energy}

\begin{minipage}{18pc}
\includegraphics[width=18pc]{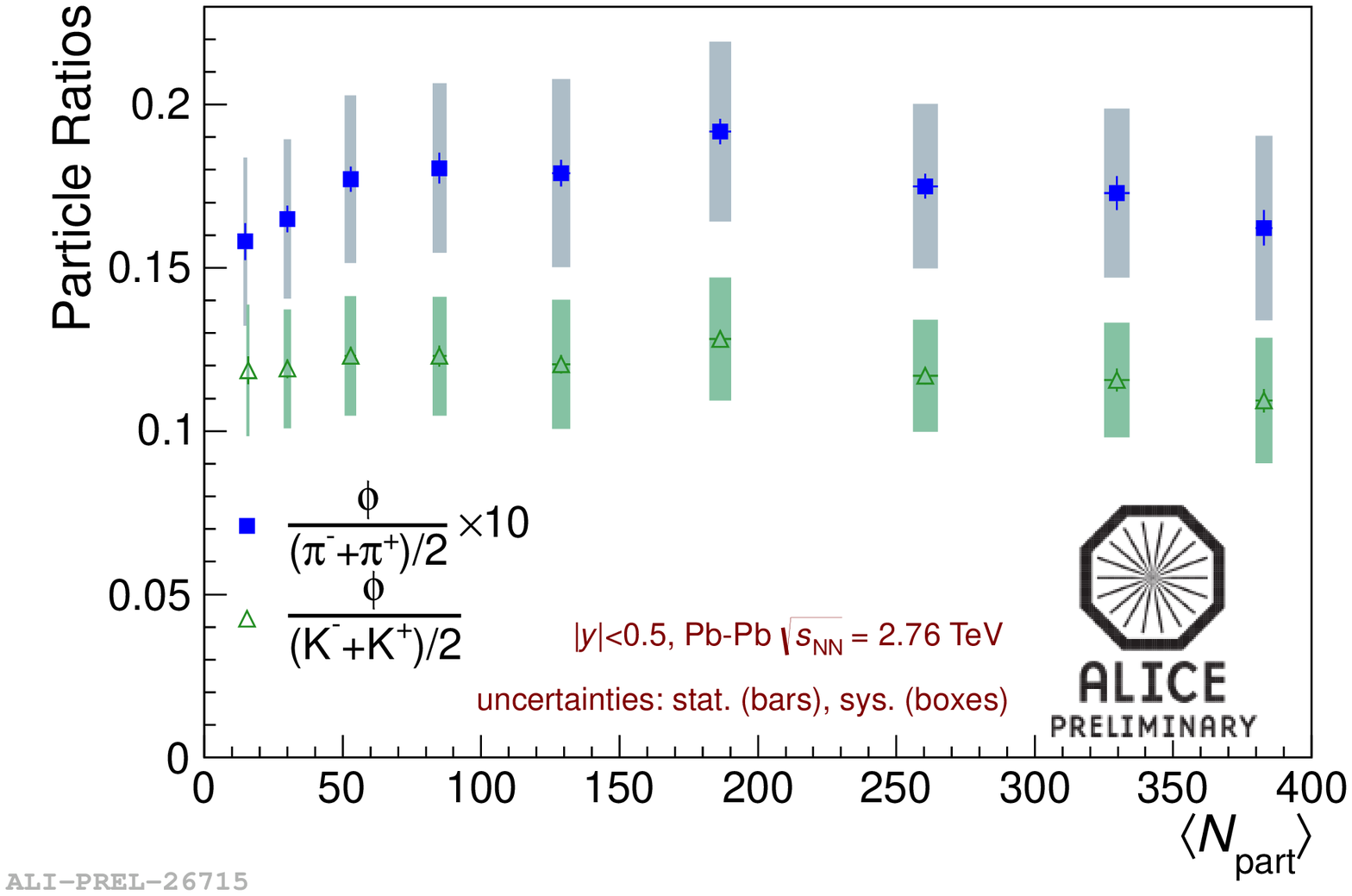}
\end{minipage}
\begin{minipage}{0pc}
\hspace{-15.5pc}\vspace{9pc}
(a)
\end{minipage}
\begin{minipage}{18.5pc}
\includegraphics[width=18.5pc]{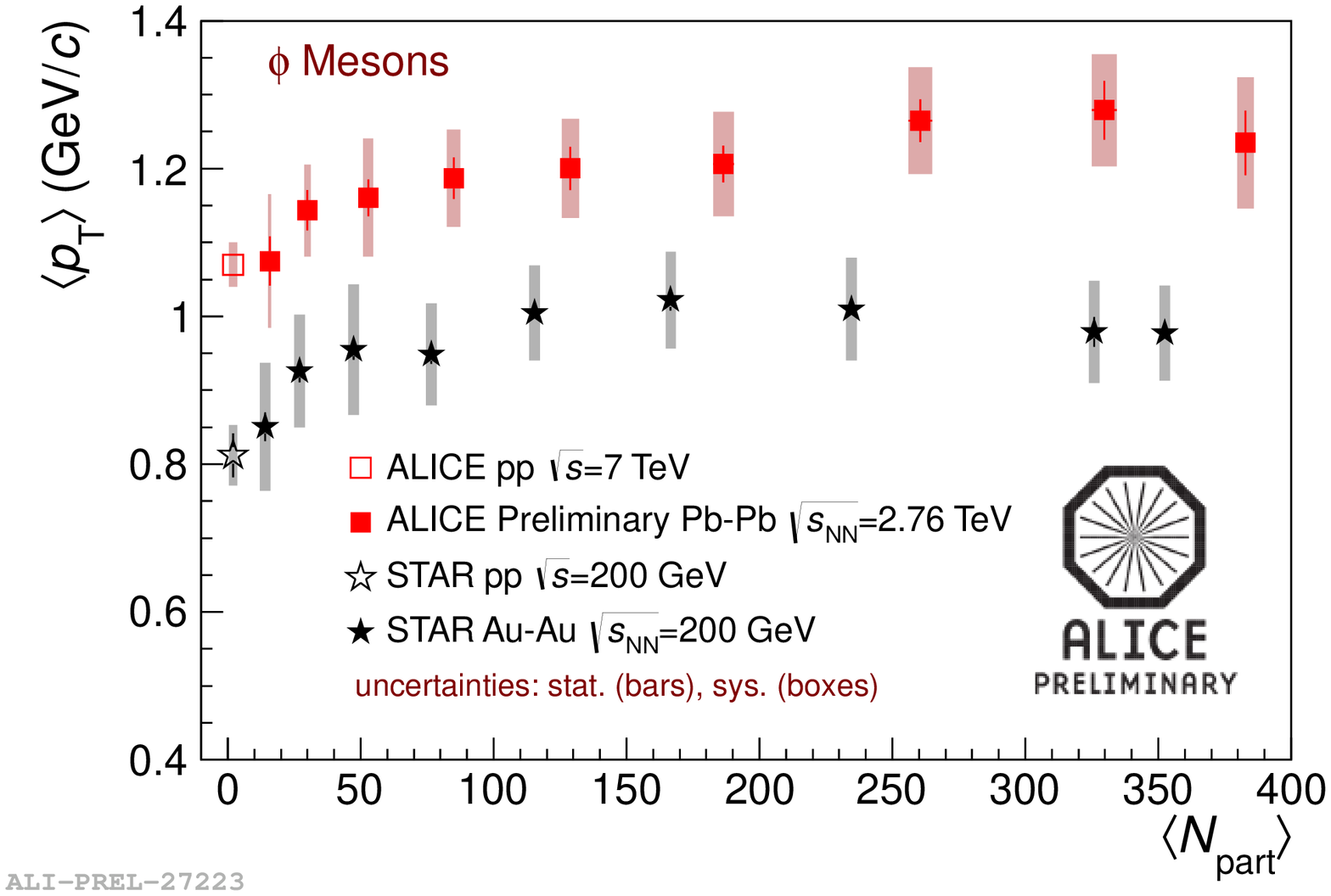}
\end{minipage}
\begin{minipage}{0pc}
\hspace{-15.75pc}\vspace{-5pc}
(b)
\end{minipage}
\caption{(a): The $\phi/\langle\pi^{\pm}\rangle$ and $\phi/\langle \mathrm{K}^{\pm}\rangle$ ratios as functions of \npart in \pb collisions at \stwo.  (b):  Mean transverse momentum for \ph mesons measured using the ALICE detector (in \pb collisions at \stwo and pp collisions at \ssvn~\cite{ALICE_KstarPhi_7TeV}) and the STAR detector (in pp and Au--Au collisions at \mbox{$\sqrt{s_{\mathrm{NN}}}=200$ GeV}~\cite{STAR_phi_2009,STAR_phi_200GeV_2005}).}
\label{fig:ratios_mpt:npart}
\end{figure}

The total \ptt-integrated yields \dndy of \ph mesons are calculated by integrating the spectra and using the fits to estimate the \ph yields for \mbox{$\ptt < 0.5$ \gvc} and \mbox{$\ptt > 5$ \gvc} (the \ph yield in the extrapolation regions is $\approx 15\%$ of the total yield).  The total \ptt-integrated yields of charged pions and kaons have been measured in \pb collisions at \stwo using the ALICE detector; these integrated yields are used to calculate the $\phi/\pi$ and $\phi/\mathrm{K}$ ratios.  Figure~\ref{fig:ratios_mpt:phi_ratios_energy} shows the $\phi/\pi$ and $\phi/\mathrm{K}$ ratios as functions of collision energy from SPS energies to LHC energies.  Figure~\ref{fig:ratios_mpt:npart}a shows the \mbox{$\phi/\langle\pi^{\pm}\rangle=\phi/\tfrac{1}{2}(\pi^{-}+\pi^{+})$} and \mbox{$\phi/\langle\mathrm{K}^{\pm}\rangle=\phi/\tfrac{1}{2}(\mathrm{K}^{-}+\mathrm{K}^{+})$} ratios as functions of \npart (the mean number of nucleons participating in collisions for a given centrality bin).

In \pb collisions, the $\phi/\langle\pi^{\pm}\rangle$ and $\phi/\langle\mathrm{K}^{\pm}\rangle$ ratios do not depend on collisions centrality.  In hadron transport models such as RQMD and UrQMD~\cite{RQMD,UrQMD,RQMD_UrQMD_2006}, where the dominant production mechanism for \ph mesons is kaon coalescence $(\mathrm{K\bar{K}}\rightarrow\phi\mathrm{X})$, the $\phi/\mathrm{K}$ ratio is predicted to be larger for more central collisions (increasing roughly linearly with \npart)~\cite{Mohanty_Xu,STAR_phi_2009}.  The $\phi/\langle\mathrm{K}^{\pm}\rangle$ ratio observed in \pb collisions at \stwo does not increase for more central events, suggesting that \ph mesons are not produced dominantly through kaon coalescence.  Similar behavior was observed at RHIC~\cite{STAR_phi_200GeV_2005,STAR_phi_2009}.  The $\phi/\mathrm{K}$ ratio does not exhibit a clear dependence on collision system or energy from RHIC to LHC energies.  The $\phi/\mathrm{K}^{-}$ ratio in \pb collisions at \stwo is consistent with the thermal model prediction~\cite{AndronicQM2011}.

The $\phi/\pi$ ratios measured in pp and \pb collisions at LHC energies are consistent with each other; at RHIC energies, the $\phi/\pi$ ratios measured using the PHENIX detector are also consistent in pp and Au--Au collisions, but the ratio measured using the STAR detector is higher for Au--Au collisions.  The $\phi/\pi^{-}$ ratio in \pb collisions at \stwo is below the thermal model prediction.

The mean transverse momentum \mpt of \ph mesons was calculated based on the data points (with the fits used to account for the low- and high-\ptt extrapolation regions).  Figure~\ref{fig:ratios_mpt:npart}b shows \mpt for the two resonances for pp and nucleus-nucleus collisions at RHIC and LHC energies.  For \pb collisions at \stwo, there may be a weak centrality dependence in \mpt, although the \mpt values in central and peripheral collisions are still consistent within uncertainties.  The \mpt values at LHC energies are higher than the observed \mpt at RHIC energies.  The \mpt values in peripheral \pb collisions at \stwo are consistent with the values measured in pp collisions at \ssvn.


\section{Jet-Resonance Correlations}
\label{sec:correlations}

\subsection{Method}
\label{sec:correlations:method}

\begin{figure}
\includegraphics[width=21pc]{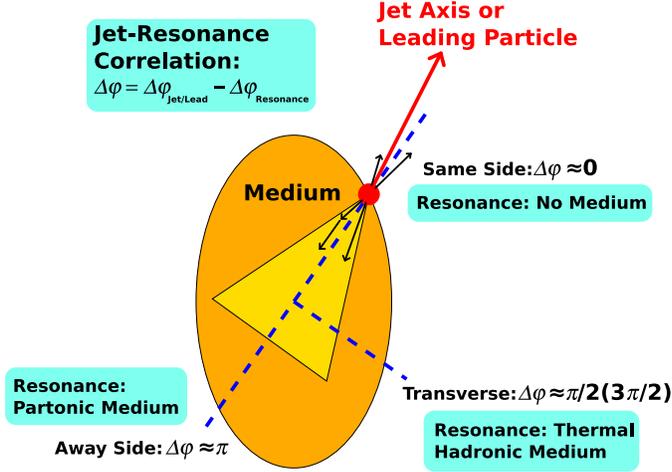}
\hspace{1pc}
\begin{minipage}[b]{14pc}
\caption{Diagram of the jet-resonance (or hadron-resonance) correlation method.  Resonances at high \ptt ($\ptt>2$ \gvc) on the away side from the trigger axis have a greater probability of interacting with the chirally restored medium.  This illustration is adapted from~\cite{Markert_corr}.}
\label{fig:correlations:diagram}
\end{minipage}
\end{figure}

The jet-resonance correlation method has been proposed~\cite{Markert_corr} as a way to preferentially select for resonances that interacted with the partonic medium or the mixed phase (and which would therefore be more likely to exhibit the signatures of chiral symmetry restoration).  While the ideal implementation of this method involves full jet reconstruction for use as a trigger, the current results use high-\ptt hadrons as triggers (serving as proxies for jets).  The difference in azimuthal angle \Dphi between a resonance and the trigger hadron is computed.  As shown in Figure~\ref{fig:correlations:diagram}, a nucleus-nucleus collision can be divided into three regions in \Dphi.  For the ``near side," ($\Dphi\approx 0$), both the jet and the resonance should be surface-biased.  It is argued that the resonance has a lower probability of interacting with the medium and therefore a lower probability of exhibiting a mass shift or width broadening.  In the ``transverse region," ($\Dphi\approx\pi/2$) resonances are expected to come predominantly from thermal production in the hadronic medium, meaning the resonances would be less likely to interact with the chirally restored medium.  In contrast, on the ``away side" from the (surface-biased) jet, for $\Dphi\approx\pi$, the probability that the resonance interacted with the medium is expected to be greater.  At low \ptt, the resonance signal (even on the away side) would be affected by interactions in the hadronic phase (re-scattering and regeneration).  At high \ptt on the away side, the resonances would be more likely both to interact with the medium when chiral symmetry was restored and to have decay products with enough momentum to escape the hadronic phase with minimal interaction.  Therefore, it was proposed that resonances with high transverse momentum ($\ptt>2$ \gvc) on the away side, opposite a trigger jet or hadron, would be more likely to exhibit a shift in mass or an increase in width, expected signatures of chiral symmetry restoration.

\subsection{Results}
\label{sec:correlations:results}

\begin{figure}
\begin{minipage}{16pc}
\includegraphics[width=16pc]{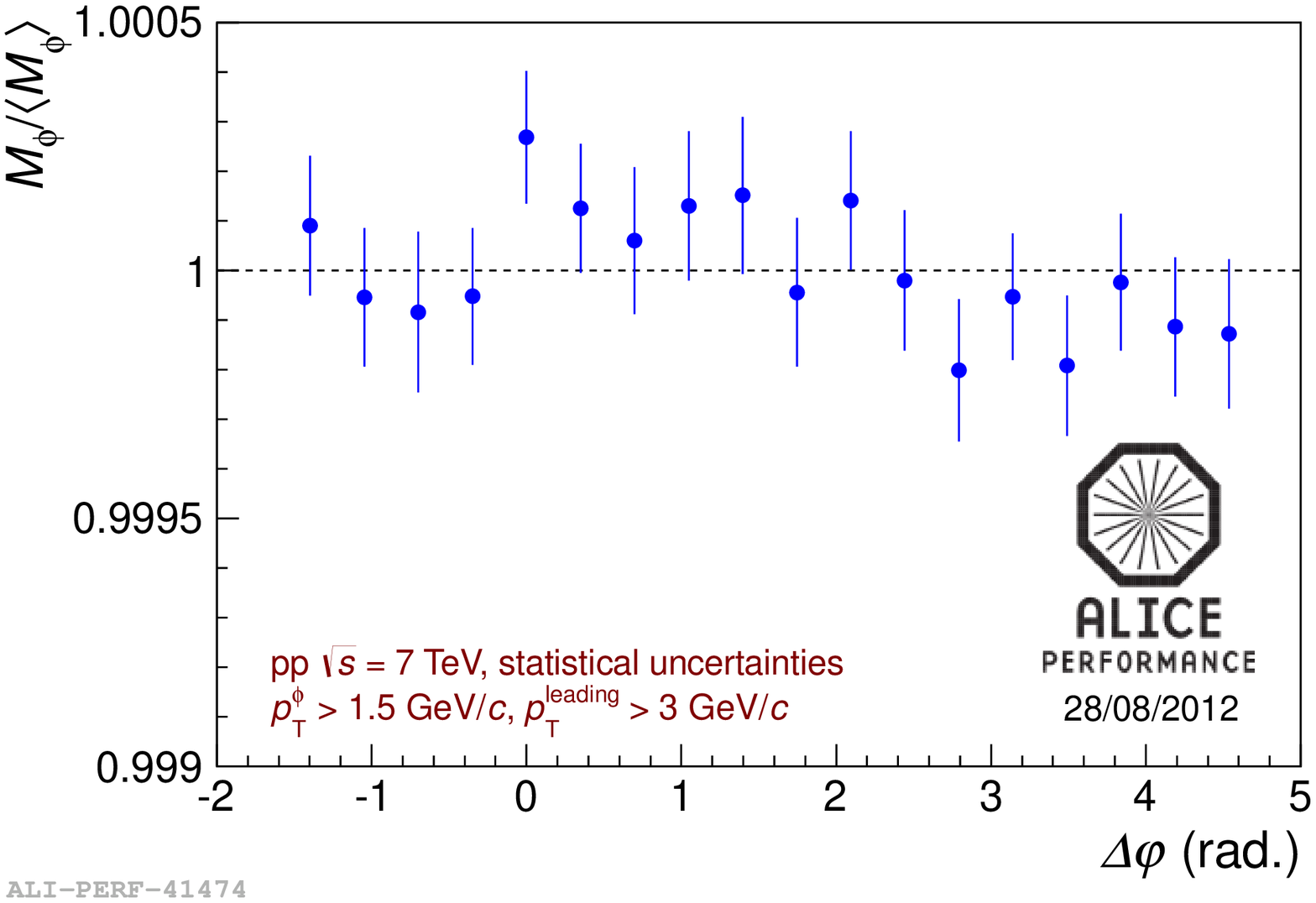}
\end{minipage}
\begin{minipage}{0pc}
\hspace{-13pc}\vspace{8.5pc}
(a)
\end{minipage}
\begin{minipage}{16pc}
\includegraphics[width=16pc]{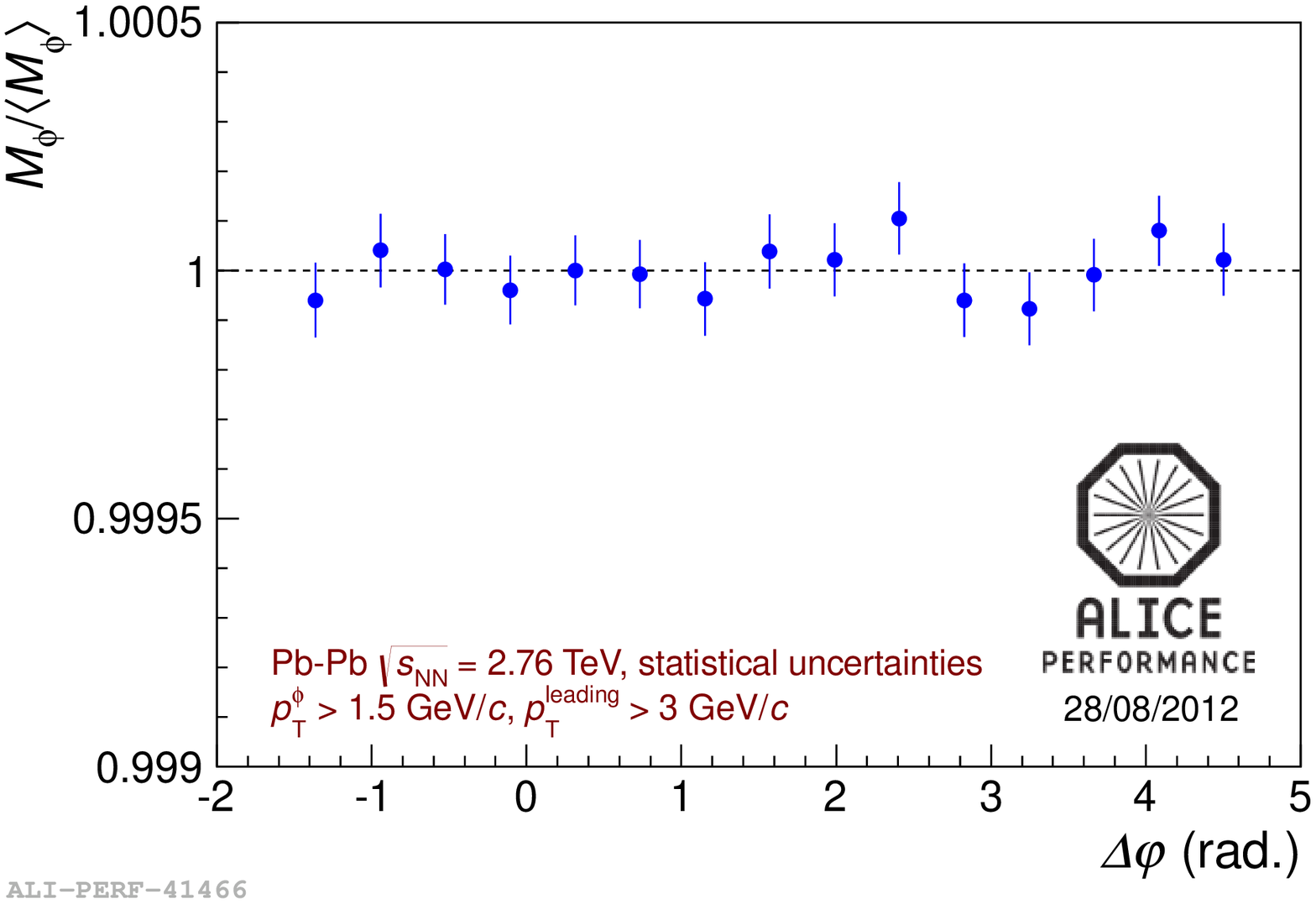}
\end{minipage}
\begin{minipage}{0pc}
\hspace{-13pc}\vspace{8.5pc}
(b)
\end{minipage}

\begin{minipage}{16pc}
\includegraphics[width=16pc]{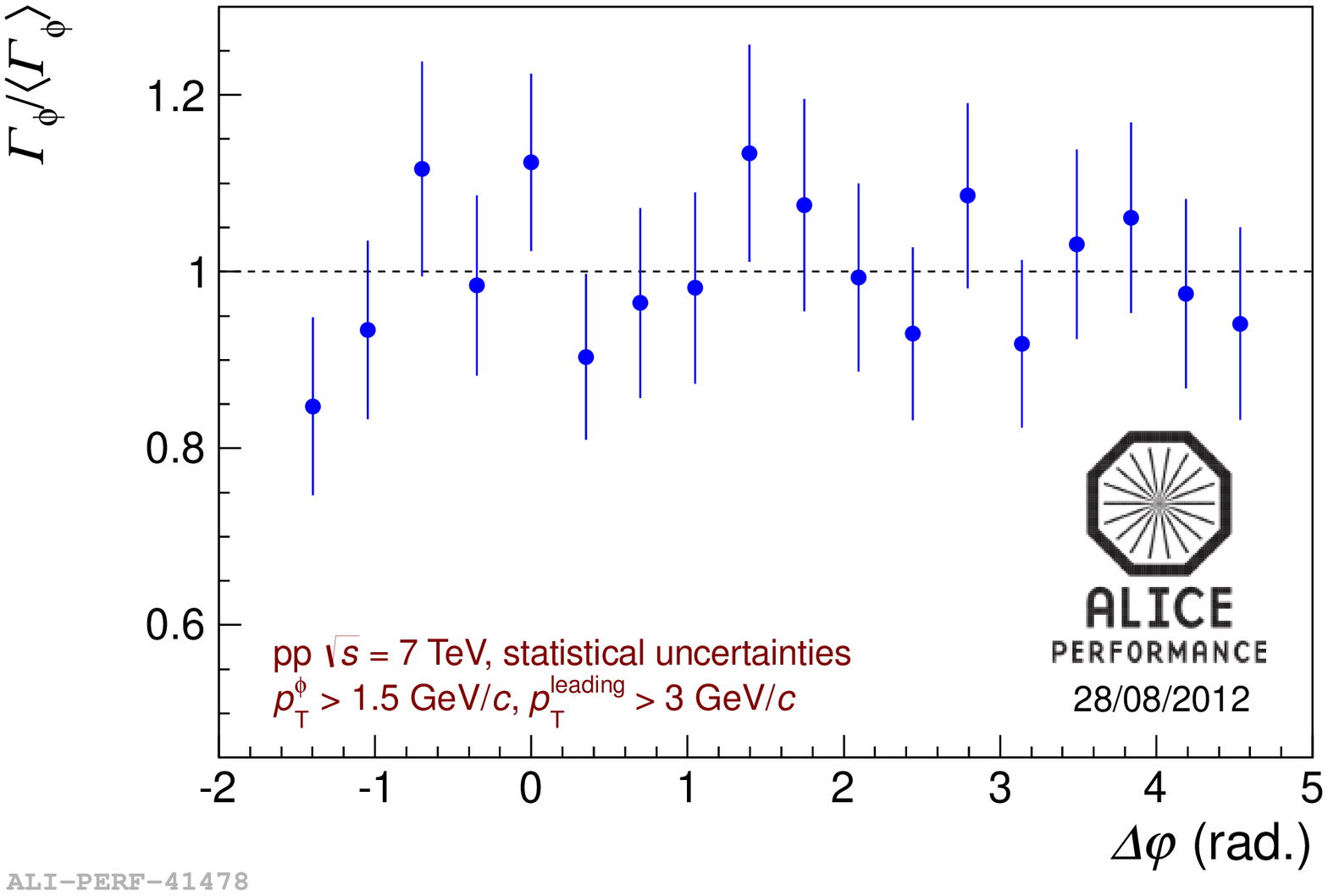}
\end{minipage}
\begin{minipage}{0pc}
\hspace{-13pc}\vspace{8.5pc}
(c)
\end{minipage}
\begin{minipage}{16pc}
\includegraphics[width=16pc]{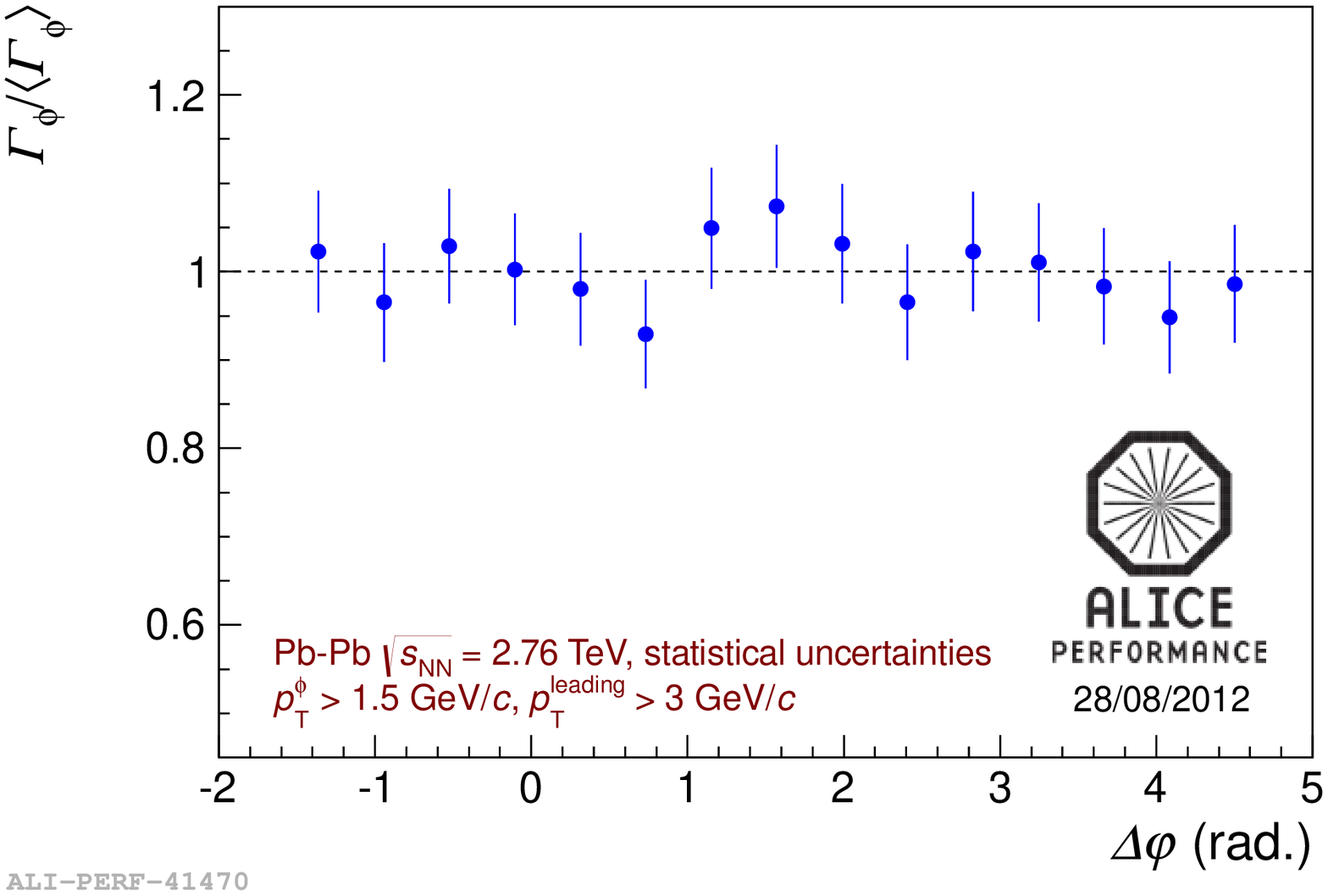}
\end{minipage}
\begin{minipage}{0pc}
\hspace{-13pc}\vspace{8.5pc}
(d)
\end{minipage}
\caption{Mass (a, b) and width (c, d) of \ph mesons vs. \Dphi, the angular correlation with a leading trigger hadron.  (a, c): pp collisions at \ssvn.  (b, d): \pb collisions at \stwo.  The leading trigger hadrons and the \ph mesons have transverse-momentum lower limits of \mbox{$p_{\mathrm{T}}^{\mathrm{leading}}>3$ \gvc} and \mbox{$p_{\mathrm{T}}^{\phi}>1.5$ \gvc}.  The mass and width values have been divided by the mean of the measured values in each plot.} 
\label{fig:correlations:results}
\end{figure}

Using the ALICE detector, \Dphi has been measured~\cite{Markert_SQM2011}, using leading trigger hadrons with \mbox{$p_{\mathrm{T}}^{\mathrm{leading}}>3$ \gvc} and \ph mesons with \mbox{$p_{\mathrm{T}}^{\phi}>1.5$ \gvc}.  Figure~\ref{fig:correlations:results} shows the mass (upper row) and width (lower row) of \ph mesons (divided by the mean of the measured values) as functions of \Dphi in pp collisions at \ssvn (left column) and \pb collisions at \stwo (right column).  The behavior of this distribution is similar for pp and \pb collisions.  Furthermore, there is no clear difference in the mass and width measured in the away side ($\Dphi\approx\pi$) in \pb collisions in comparison to the near side ($\Dphi\approx 0$).


\section{Conclusion}
\label{sec:conclusion}

The spectra (for $\ptt<5$~\gvc), total yields, mean transverse momenta, mass, and width of the \ph meson and the mass and width of the \ks meson have been measured in \pb collisions at \stwo using the ALICE detector.  No centrality dependence has been observed in the masses or widths of either resonance.  The \ks mass values are not consistent with the nominal value at low \ptt, but are consistent with the mass measured in pp collisions (indicating that the observed deviation is likely an instrumental effect).  The \ks width is consistent with the nominal value.  The mass of the \ph meson is within 0.5 \mvcc of the nominal value, while the width is within 1-2 \mvcc of its nominal value; similar deviations are observed when simulated \ph mesons are analyzed using the same methods.  The $\phi/\pi$ and $\phi/\mathrm{K}$ ratios in \pb collisions do not depend on collision centrality, indicating that \ph mesons are not produced predominantly through kaon coalescence.  These ratios are consistent with the values measured in pp collisions at \mbox{$\sqrt{s}=900$ GeV} and \ssvn.  The \mpt values for \ph mesons at LHC energies are higher than the values measured at RHIC energies.  The angular correlations \Dphi between leading trigger hadrons and \ph mesons have been measured in pp and \pb collisions, but no difference in the resonance mass or width between the near and away sides (relative to the trigger axis) has been observed.  However, it may be necessary to increase the transverse-momentum limits on the trigger particles and the \ph mesons in order to obtain a sample of resonances that exhibit the signatures of chiral symmetry restoration.

\section*{References}
\bibliography{NN2012_Knospe}

\end{document}